\documentclass[aps,pre,preprint,onecolumn,amsmath,amssymb]{revtex4}
\usepackage{makecell}
\usepackage{bbm}
\usepackage{mathrsfs}
\usepackage{subfigure}
\usepackage{multirow} 
\usepackage{bigstrut} %
\usepackage{amsfonts}
\usepackage{color}
\usepackage{graphicx}
\usepackage{boxedminipage}

     \usepackage{hyperref} 
 \hypersetup{
 colorlinks=true,
 linkcolor=blue,
 urlcolor=blue,
 }

\newcommand{\be}{\begin{equation}}
\newcommand{\ee}{\end{equation}}
\newcommand{\ba}{\begin{eqnarray}}
\newcommand{\ea}{\end{eqnarray}}

\usepackage{hyperref}
\usepackage{tikz}

\begin{document}

\title{Capturing the Demon in Szilard's Engine: \\Partition Reset as Information Erasure}
\author{Xiangjun Xing$^{1,2,3}$}
\email{xxing@sjtu.edu.cn}
\address{$^1$Shanghai Jiao Tong University, Shanghai 200240, China \\
$^2$T.D. Lee Institute, Shanghai Jiao Tong University, Shanghai 200240, China\\
$^3$Shanghai Research Center for Quantum Sciences, Shanghai 201315, China}
\date{\today}

\begin{abstract}
In Szilard’s engine, a demon inserts a partition into a one-particle gas box and moves it towards the empty side, extracting work from a thermal bath, which seemingly violates the second law. Bennett modified this setup with a fixed partition and two movable pistons, resolving the paradox via memory erasure. In this work, we revisit Szilard's setup, showing the movable partition is an information-bearing degree of freedom whose reset incurs an entropy cost. This resolves the entropy paradox while eliminating the necessity to introduce any demon. Our re-analysis of Szilard's engine  illustrates a general methodology for study of information thermodynamics. 
\end{abstract}

\maketitle

\section{Introduction}

In 1929, as an effort to resolve the puzzle raised by Maxwell's demon, Leo Szilard published a seminal paper~\cite{Szilard1929} titled ``{\"U}ber die Entropieverminderung in einem thermodynamischen System bei Eingriffen intelligenter Wesen'' (English translation: On the Decrease of Entropy in a Thermodynamic System by the Intervention of Intelligent Beings), where he introduced a thought experiment that illuminated the interplay between information and thermodynamics. In this thought experiment, later widely known as {\em Szilard's engine}, a demon inserts a partition at the midpoint of a box containing a one-particle ideal gas which is in thermal contact with a bath at temperature \( T \).  By measuring the particle’s position---left or right of the partition with equal probability---the demon moves the partition toward the empty side to initiate a quasi-static expansion of the gas, which extracts mechanical work at seemingly no other cost.  This apparently violates the second law of thermodynamics.  Szilard proposed that the demon's measurement must incur an entropy production which is sufficient to restore consistency with the second law.  

Szilard’s paper sparked decades of debate on the thermodynamic role of information, and also drew numerous critiques, primarily focused on the subjective nature of the demon’s measurement and the questionable link between information and entropy~\cite{footnote-criticism-Szilard}.  In 1961, Rolf Landauer~\cite{Landauer1961} introduced his principle, establishing that logically irreversible operations, such as erasing a bit by merging states, dissipate at least \( k_B T \ln 2 \) as heat.  This principle provides a physical basis for the link between information and thermodynamic entropy.  In 1982, Charles Bennett applied this principle to a modified Szilard engine with a fixed partition and two movable pistons, resolving the paradox by attributing the missing entropy cost to memory erasure \cite{Bennett1982}.  The combination of theories of Landauer and of Bennett, popularly known as the {\em Landauer-Bennett Thesis}~\cite{Landauer1961,Bennett1982,Bennett1973,Bennett1988,Landauer1996,Landauer1996-2,Bennett-2002}, has been widely perceived as a proper resolution to the paradox raised by Szilard's engine and other toy models of Maxwell's demon.   The Landauer-Bennett Thesis has been confirmed in a variety of experimental systems~\cite{Toyabe2010,Berut2012}, where the demon's role is played by modern computerized lab devices.  

There is however an important difference between Szilard's original thought experiment and Bennett's modification, which has never been properly addressed.  In Bennett's modified setup, the partition is inserted in the middle and remains fixed, whereas two side walls of the gas box are replaced by two movable pistons.  The demon moves the piston in the empty side to drive the gas expansion, such that all mechanical devices, including the partition and two pistons, return to their initial positions after the expansion and the extraction of work. The only change is in the demon's memory.  By contrast, in Szilard's original setup, the partition both serves as divider of the gas box and as a piston that drives the gas expansion.  At the end of the expansion, the partition is at either side of the box, which breaks the left/right symmetry.  Even though Szilard believed that the partition can be reset to its original position with no other consequence, this is not true in the light of Landauer's principle.  Reset of the partition is equivalent to erasure of information encoded it encodes, which does incur a thermodynamic cost.  This is an essential loophole in Szilard's original reasoning which has been long overlooked.   It also obstructs a straightforward application of Landauer-Bennett Thesis to Szilard's original thought experiment---even if the demon's memory is erased, the partition has not been restored.   It is the purpose of this work to re-analyze Szilard's original thought experiment using Landauer's principle, and to demonstrate that once the reset cost of the partition is taken into account, there is no longer need to introduce any demon---the partition is the demon.  

The remainder of this article is organized as follows. In Section~II we review carefully the original setup of Szilard and Bennett's modification, together with Landauer-Bennett thesis which resolves the entropy paradox in Bennett's setup.  In Section III we treat the movable partition in Szilard's setup as an information-bearing variable and show that its entropy cost of resetting resolves the entropy paradox. We further supply entropy balance sheets both for Szilard's setup and for Bennett's setup, verifying the reversible nature of both processes.  In Section IV we conclude with some broader perspectives.

\section{Szilard's Engine and Bennett's Modification}

\subsection{Szilard's original setup}

\begin{figure}[t!]
    \centering
     \includegraphics[width=5.5in]{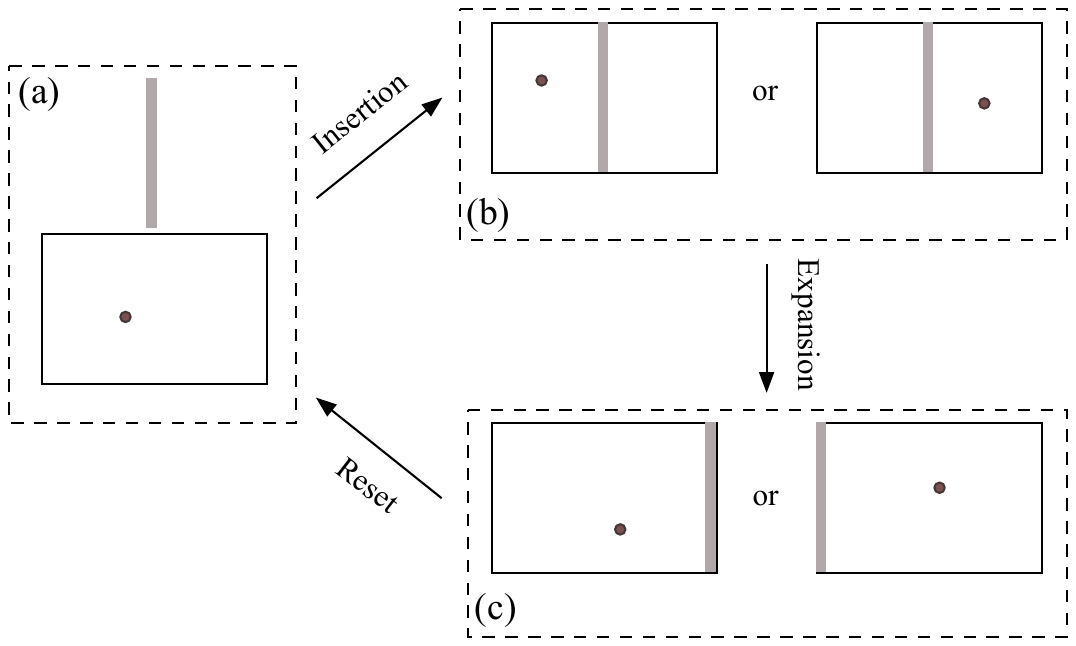}   
    \caption{Szilard's cycle in three states (a), (b), and (c) with transitions.  Insertion (a) $\to$ (b). At (b), the demon measures the location of the particle in order to choose the direction of expansion.   Expansion (b) $\to$ (c): extraction of work $T$, and recording of the past position of the particle before expansion.    Reset (c) $\to$ (a): resetting the partition to the initial position. Without capturing the thermodynamic cost of resetting the partition, Szilard argued that there must be an entropy production associated with the measurement. }
    \label{fig:partition-Demon-schematics}
\end{figure}

Let us follow closely the original setup of Szilard~\cite{Szilard1929} with some non-essential simplifications.  Consider a single-particle ideal gas in a box of volume \(V\) and in thermal contact with a bath at temperature \(T\), as schematized in Fig.~\ref{fig:partition-Demon-schematics}(a).  A demon inserts a partition (German: {\em Zwischenwand}) in the center of the box~\footnote{In fact, Szilard assumed that the partition is inserted off the center, which makes the discussion slightly more complicated. }, confining the particle to the left or right side with equal probability (50\%), arriving at state (b) in Fig.~\ref{fig:partition-Demon-schematics}.  The demon then measures the particle location (L or R) and initiates a quasi-static expansion by moving the partition towards the empty side.  As Szilard said in his paper: ``Diese Zwischenwand bildet einen Stempel, der sich im Zyllinder nach...'' (This
partition forms a piston that can be moved...).   Hence it is clear that, in Szilard's paper,  Zwischenwand (partition) and Stempel (piston) refer to the same thing that both divide the box and drive the expansion.  At the end of the expansion, the partition arrives at the left/right end of the box and the volume of the gas restores to $V$, arriving at state (c) in Fig.~\ref{fig:partition-Demon-schematics}.  The work done by the partition during the expansion stage is 
\begin{equation}
W = \int_{V/2}^{V} P \, dV 
=  T \int_{V/2}^{V} \frac{1}{V'} \, dV' 
= T \log_2  \left( \frac{V} {V/2} \right) = T,
\label{eq:work}
\end{equation} 
where used was the ideal gas law.  This extracted work of $T$ corresponds to one bit of entropy reduction in the bath, consistent with the later Landauer bound. Throughout this article, we set the Boltzmann constant $k_{\rm B} = 1$ and use base-2 logarithms to define entropy. Thus, we express temperature in energy units and entropy in bits, a convention that greatly simplifies discussions of information thermodynamics.  The work (\ref{eq:work}) comes solely from the internal energy of the bath.  Therefore, as consequences of the expansion, the energy of the heat bath decreases by \(  T \), and the entropy of the bath decreases by \(\Delta S_{\text{bath}} = -  1 \, {\rm bit}\).  Szilard then says~\cite{Szilard1929}  ``$\cdots$ und der Stempel wird alsdann herausgezogen. Der Vorgang kann beliebig oft wiederholt werden. '' (English translation: $\cdots$ the piston is then withdrawn. The process can be repeated as many times as desired, see page 112 of Ref.~\cite{LeffRex2003}).  If the withdrawal, i.e., reset, of the piston is free of any cost, the cycle would lead to an apparent violation of the second law.  (However, the reset of the piston is not free.)

To save the second law, Szilard argued: ``Es ist nun naheliegend, anzunehmen, dass die Vornahme einer Messung prinzipiell mit einer ganz bestimmten mittleren Entropieerzeugung verbunden ist, und dass dadurch der Einklang mit dem zweiten Hauptsatz wieder hergestellt wird; grosser durfte die bei der Messung entstehende Entropiemenge freilich immer sein, nicht aber kleiner.'' (English translation: It is now natural to assume that the performance of a measurement is, in principle, associated with a specific average entropy production, and that this restores consistency with the second law; however, the amount of entropy generated during the measurement must always be greater, never less.)  Since the physical process of measurement remains unspecified, Szilard's statement about its entropy cost remained a speculation, which has been criticized in numerous later works~\cite{footnote-criticism-Szilard}.

\begin{figure}[t!]
\centering
     \includegraphics[width=4in]{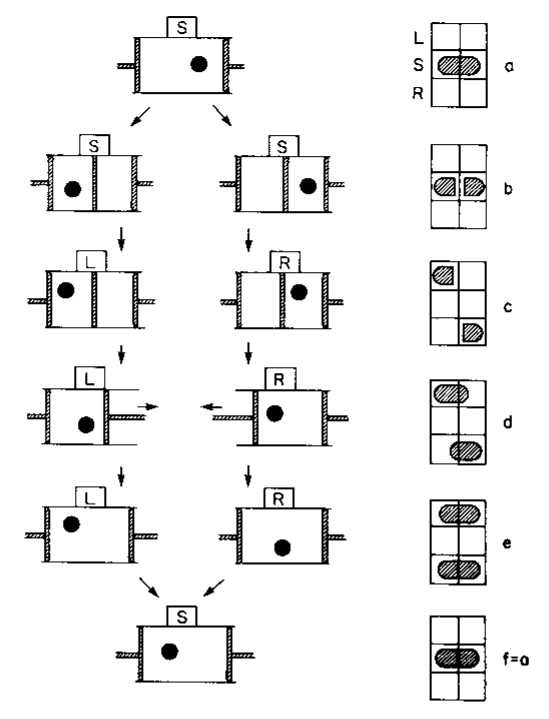}   
\caption{Bennett's modification of Szilard's cycle, extracted from \cite{Bennett1982}. Labels a, b, c, d, e, f in the figure correspond to (a), (b), (c), (d), (e), (f) in this caption.  
The left half of the figure shows the states of the gas, whereas the right half shows the states of demon's memory. Note that the gas is confined by two movable pistons.  (a): (the second row) the initial state. (b): a partition is inserted at the midpoint. (c): the demon measures the particle and records the result. The piston in the empty side then moves to the center, and the partition can be withdrawn (not shown).   (d): the center piston expands towards its initial position.  (e): the piston returns to its initial position, so that all mechanical components are restored, and an amount of work equal to $T$ is extracted.  The only changes are the demon's memory and the bath entropy.  (f): the demon's memory is erased, which incurs a heat dissipation of $T$. Everything returns to its initial state.  }
    \label{fig:Bennett-schematics}
\end{figure}

\subsection{Bennett's modification and Landauer-Bennett Thesis}

In 1961, Landauer~\cite{Landauer1961} proposed that logically irreversible operations, which erase information, necessarily dissipate energy as heat to the environment. This idea, later termed {\em Landauer's principle}, provides the first link between information processing and thermodynamics.  Bennett~\cite{Bennett1982} reformulated Szilard's setup.  He fixed the partition at the center, but allowed the left and right walls of the gas box to move as pistons, as illustrated in Fig.~\ref{fig:Bennett-schematics}.  The modified setup is such that, when the expansion is done, both the partition and the pistons are restored to their initial positions, whilst the bath energy reduces by $T$.   If the internal state of the demon has not changed, work of $T$ would have been extracted from the bath with no cost, which violates the second law.  
Bennett then argued that to decide the direction of expansion, the demon must measure the location of the particle and record the result in its internal memory, which leads to entropy increase of the memory by one bit.  To make the process fully cyclic, one would have to erase the information stored in the memory.  This is a logically irreversible operation, which, according to Landauer's principle, requires dissipation of energy at least $T$ back to the bath. The net change of total entropy is then no less than zero, saving the second law.   The combination of Landauer's and Bennett's theories soon became widely accepted as the foundation of information-thermodynamics, known as the {\em Landauer-Bennett Thesis}.

Bennett's argument adequately resolves the entropy paradox in his modified setup, becuase the total process including erasure of the demon's memory is truly cyclic.  This argument however does not address the loophole in Szilard's original reasoning---the neglect of the entropy cost for the reset of the partition in his setup.

\section{Revisiting Szilard's Original Setup}
Let us go back to Szilard's original setup.    At the end of the expansion, the position of the partition encodes the direction of expansion, and therefore is an {\em information-bearing variable}, which is conceptually equivalent to a demon's memory.  Resetting of the partition leads to merging of two inequivalent logical states, and therefore, by Landauer's principle, must dissipate heat.   It is then clear that there is really no need to introduce demon's memory in Szilard's original setup.  In fact, the entire demon can be removed from the story. Consider state Fig.~\ref{fig:partition-Demon-schematics}(b),  where the partition is inserted at the center.  Via some appropriate mechanical setup, we can make the partition sense the pressure difference caused by the one-molecule gas.  This pressure difference may autonomously trigger a reversible expansion.  This expansion simply reflects the system's spontaneous tendency to converge to an entropy-maximizing equilibrium, consistent with thermodynamic principles.  The direction of the expansion, as well as the location of the partition, effectively record the particle's position (left or right).   Therefore, the partition, which drives the expansion, also acts as a memory register.  If the particle is on the left, the partition expands rightward to the right wall; if on the right, leftward to the left wall, see Fig.~\ref{fig:partition-Demon-schematics} (c).   After the expansion, we must reset the partition to its initial position to make the process truly cyclic.  This reset is logically irreversible and requires at least work of \(  T \) to be dissipated into the bath.  The full process (a) $\rightarrow$ (b) $\rightarrow $ (c) $\rightarrow $ (a), as shown in Fig.~\ref{fig:partition-Demon-schematics} is cyclic, and there is no change in the total entropy of the universe.   

Our above analysis of Szilard's original setup fully reconciles the idea of memory register with the ideas of many Landauer-Bennett's objectors~\cite{footnote-criticism-Szilard} that no measurement/memory register is needed before work extraction, and greatly simplifies the logical structure of information-thermodynamics interplay.   According to this {\em Partition-Demon Thesis}, the process of expansion is also the process of measurement/feedback: they are the one and same thing, because the partition's motion both reveals and responds to the particle’s location simultaneously. 
 This Partition-Demon Thesis eliminates the need for any fictitious demon - {\em the partition is the demon}.   In other words, what we have realized is an autonomous version of Maxwell's demon, where the full process can be realized without interference from any intelligent being. 

We emphasize that our analysis does not contradict that of Bennett, as they apply to different processes.

\subsection{Entropy balance sheets}  
To make the information-thermodynamics more transparent, we supply a detailed study of entropy balance  both for Szilard's original setup and for Bennett's modification.  We assume that in both setups, the expansion and reset are both quasistatic and frictionless, so that the overall process is thermodynamically reversible, with zero change in the total entropy.  

 In Szilard's original setup as shown in Fig.~\ref{fig:partition-Demon-schematics}, we use $S(G_t), S(P_t)$ to denote the entropies of the gas and the partition at time $t$, and use $I(G_t;P_t)$ to denote their mutual information.  Their joint entropy is then:
\ba
S(G_t,P_t) = S(G_t)  - I(G_t;P_t) + S(P_t). 
\ea 
 Note that $S(G_t)$ and $S(P_t)$ are both {\em unconditional entropies}.  The initial entropy of the gas in (a) is $S(G_a) = \log V/\lambda_T^3$, where $\lambda_T$ is the thermal de Broglie wavelength.  If we knew that the particle is on the left half box after insertion of the partition, the conditional entropy of the gas given the information would be $ \log (V/2 \lambda_T^3) = \log (V/\lambda_T^3) -1$.  But that conditional entropy is irrelevant in Partition-Demon Thesis --- the partition does not measure the particle right after the insertion. Instead, we infer that information after the expansion is done.  

We use $S(B_t)$ to denote the thermodynamic entropy of the bath. We assume that there is no correlation between the bath and the gas/partition.  Hence the total entropy of the universe is
\ba
S_{\rm tot} &=& S(G_t,P_t)  + S(B_t)
\nonumber\\
&=& S(G_t)  - I(G_t;P_t) + S(P_t) + S(B_t). 
\ea
We use $S_0$ to denote the value of bath entropy in the initial state (a) of the cycle.  The breakdown of the total entropy into various components in each of states (a), (b), (c) is listed in Table \ref{tab:partition-Demon-entropy}.  Note that the total entropy remains the same in all states, consistent with the reversible nature of ideal Szilard's cycle. 

{It is important to note that in state (c), there is no mutual correlation between the partition and the gas---their mutual information $I(G_c;P_c)$ is identically zero.  This does not contradict the fact that there is a perfect mutual correlation between the partition $P_c$ in state $c$ and the state of the gas $G_b$ in state $b$: $I(G_b; P_c) = 1$.  That is, the correlation is {\em retrodictive} (final partition vs. past gas state), not simultaneous.  The final position of the partition encodes the  direction of expansion and the whereabouts of the gas particle immediately after the insertion. As we emphasized above, this mechanism of memory register is autonomous---it does not need any intervention from a demon.}

\begin{table}[t!]
    \centering
    \begin{tabular}{|c|c|c|c|c|c|}
        \hline
        State & $S(G)$ & $I(G;P)$ & $S(P)$ & $S(B)$ & $S_{\text{tot}}$ \\
        \hline
        (a)   & \quad $\log  (V / \lambda_T^3)$ \quad& 0 & 0 & $S_0$ &\quad $\log  (V / \lambda_T^3) + S_0$\quad \\
        (b)   &\quad $\log  (V / \lambda_T^3)$ \quad& 0 & 0 & $S_0$ &\quad $\log  (V / \lambda_T^3) + S_0$ \quad\\
        (c)   &\quad $\log  (V / \lambda_T^3)$ \quad& 0 & 1 & $S_0 - 1$ & \quad$\log  (V / \lambda_T^3) + S_0$\quad \\
        (a)   &\quad $\log  (V / \lambda_T^3)$ \quad& 0 & 0 & $S_0$ &\quad $\log  (V / \lambda_T^3) + S_0$ \quad\\
        \hline
    \end{tabular}
    \vspace{3mm}
    \caption{Entropy balance sheet for Szilard's cycle, which is illustrated in Fig.~\ref{fig:partition-Demon-schematics}.  $S(G)$, $S(P)$, and $S(B)$ are entropies of the gas, partition, and bath; $I(G;P)$ is mutual information between current gas and partition states (0 bits at (c), though partition correlates with past gas state); $S_{\text{tot}}$ is total entropy. Note that the total entropy remains invariant, showing the reversibility of Szilard's cycle. }
    \label{tab:partition-Demon-entropy}
\end{table}

In Bennett's reformulation of Szilard engine as shown in Fig.~\ref{fig:Bennett-schematics}, we use $S(D)$ to denote the entropy of the demon, and use $I(G;D)$ to denote its mutual information with the gas.  The joint entropy of the gas and the demon is then:
\ba
S(G,D) = S(G)  - I(G;D) + S(D). 
\ea 
The total entropy of the universe  is
\ba
S_{\rm tot} = S(G,D)  + S(B)
= S(G)  - I(G;D) + S(D) + S(B). 
\ea
The breakdown of the total entropy into various components in each of states (a), (b), (c), (e), (f) = (a) is listed in Table \ref{tab:Landauer-Bennett-entropy}. (The state (e) in Fig.~\ref{fig:Bennett-schematics} is a generic intermediate state that is not further analyzed here.)  Note that a mutual information between G and D is established in state (c) right after the measurement.  Note also that the total entropy remains the same in all states, consistent with the reversible nature of the ideal cycle.

\begin{table}[ht!]
    \centering
    \begin{tabular}{|c|c|c|c|c|c|}
        \hline
        State & $S(G)$ & $I(G;D)$ & $S(D)$ & $S(B)$ & $S_{\text{tot}}$ \\
        \hline
        (a)   & \quad $\log  (V / \lambda_T^3)$ \quad& 0 & 0 & $S_0$ &\quad $\log  (V / \lambda_T^3) + S_0$\quad \\
        (b)   &\quad $\log  (V / \lambda_T^3)$ \quad& 0 & 0 & $S_0$ &\quad $\log  (V / \lambda_T^3) + S_0$ \quad\\
        (c)   &\quad $\log  (V / \lambda_T^3)$ \quad& 1 & 1 & $S_0$ &\quad $\log  (V / \lambda_T^3) + S_0$ \quad\\
        (e)   &\quad $\log  (V / \lambda_T^3)$ \quad& 0 & 1 & $S_0 - 1$ & \quad$\log  (V / \lambda_T^3) + S_0$\quad \\
        (f) = (a)   &\quad $\log  (V / \lambda_T^3)$ \quad& 0 & 0 & $S_0$ &\quad $\log  (V / \lambda_T^3) + S_0$ \quad\\
        \hline
    \end{tabular}
    \vspace{3mm}
    \caption{Entropy balance sheet for Bennett's reformulation of Szilard's cycle, which is illustrated in Fig.~\ref{fig:Bennett-schematics}.  $S(D)$ is the entropy of the demon.   }
    \label{tab:Landauer-Bennett-entropy}
\end{table}

\section{Conclusion}

In this work, we revisited Szilard’s engine, long regarded as the paradigmatic link between information and thermodynamics.  While the Landauer-Bennett Thesis resolves the paradox in Bennett's modified setup by attributing the missing entropy cost to memory erasure, it does not do the same job in Szilard's original setup---The missing piece in Szilard’s analysis is not the entropy cost of measurement or abstract memory erasure, but the reset cost of the displaced partition.  Our re-analysis shows that the partition itself acts as an information-bearing variable. Its final position encodes the outcome of the expansion, and resetting it requires at least an amout of heat $T $, thereby restoring consistency with the second law. Once this is recognized, the need for an abstract demon disappears---the partition itself \emph{is} the demon.  

Our re-analysis of Szilard's engine illustrates a general methodology for the study of information thermodynamics: we must always begin with identifying the concrete information-bearing variables of a system, analyze its dynamics, and quantify its thermodynamic consequences.  From this standpoint, Bennett's analysis remains incomplete, since the demon's memory is left without a concrete physical model.  To render it concrete would require a detailed physical model of the demon’s dynamics, a task that depends on many specifics and has not been completely carried out to date. We expect that further studies along these lines will extend the general methodology to other demon models and non-equilibrium systems, consolidating the role of information-bearing variables as the foundation of the field.


The author thanks Grok 3, an AI assistant developed by xAI, for both enlightening scientific discussions and useful editorial assistance.   The author has no conflicts to disclose.

\end{document}